\documentclass[11pt]{article}

\usepackage{sectsty}
\usepackage{graphicx}
\usepackage{natbib}
\usepackage{booktabs}
\usepackage{amsmath}
\usepackage{amsfonts}
\usepackage{amssymb}
\usepackage{rotating}
\usepackage{hyperref}
\begin{document}

\newcommand{\vz}{\boldsymbol{z}}
\newcommand{\vx}{\boldsymbol{x}}
\newcommand{\vb}{\boldsymbol{b}}
\newcommand{\vs}{\boldsymbol{s}}

\title{Providing Previously Unseen Users Fair Recommendations Using Variational Autoencoders}

\author{Bjørnar Vassøy, Helge Langseth and Benjamin Kille\\ Norwegian University of Science and Technology}
\date{}

\maketitle

\begin{abstract}
  An emerging definition of fairness in machine learning requires that models are oblivious to demographic user information, e.g., a user's gender or age should not influence the model. Personalized recommender systems are particularly prone to violating this definition through their explicit user focus and user modelling. Explicit user modelling is also an aspect that makes many recommender systems incapable of providing hitherto unseen users with recommendations. We propose novel approaches for mitigating discrimination in Variational Autoencoder-based recommender systems by limiting the encoding of demographic information. The approaches are capable of, and evaluated on, providing users that are not represented in the training data with fair recommendations.
\end{abstract}

\section{Introduction}
Fairness in recommender systems is becoming a popular and diverse research field. \citet{burke_multisided_2017} formalized the multi-stakeholder nature of the recommendation setting: Producers are interested in mitigating popularity bias \citep{pop_bias} to ensure their products are given the exposure they deserve. Consumers expect to be treated similarly regardless of demographic attributes like age, race, and gender. In addition to the stakeholder perspectives, there is no single definition of what constitutes a fair recommendation. A consumer-side perspective focusing on discrimination of demographic user groups may require that similar ratings are estimated for each group \citep{kamishima_recommendation_2018}, that each group receives similar recommendations \citep{hybrid_farnadi}, that each group are equally satisfied with their recommendations \citep{beyond_2017}, or that model representations do not correlate with the user groups. This research focuses on the latter notion (\textit{Neutral Representations}) while also evaluating whether similar recommendations are given to different demographic groups (\textit{Recommendation Parity}).

A mostly unexplored perspective of fair recommender systems relates to recommending for users that were not part of the training data. A user's first impression of a platform may decide if they will continue using it. A man who aspires to be a florist and likes romance movies may feel stereotypes and be discouraged from further interaction if he is first recommended physical labour careers or action movies. Unlike contemporary fair recommender systems, our research focuses on introducing fairness in a model architecture that can recommend for all users, including users not represented in training data. We propose Variational Autoencoder (VAE) approaches that only require a list of items a user has interacted with and no pre-trained user representations. The approaches may replace contextual and item-to-item recommender systems used to onboard new users or serve as a complete recommendation platform that does not require frequent model updates. 

One goal of this research is to provide more insight into the competitiveness of VAE-based recommender systems given their limited use in recommender system research. Given this, we wish to verify if our VAE-based models can fairly process unseen users and if the same models produce state-of-the-art fair recommendations for said users. Formalized research questions are as follows:
\textbf{1.} Are VAE-based recommender systems competitive? and \textbf{2.} Can the encoded demographic information in VAE latent states be reduced when processing new users?

\section{Related Work}
Multiple methods have been proposed for filtering out sensitive information embedded in model representations and parameters. Many of these methods train adversarial models tasked with classifying sensitive attributes given representations belonging to the recommender systems. The recommender systems are then penalized for encoding sensitive information by adding additional objectives of fooling the adversarial models. This strategy has been applied directly to latent factors of factorization models \citep{resheff_privacy_2019,xu_fair_2021,wu_fairness-aware_2021}, indirectly to train attribute filters \citep{bose_compositional_2019}, and for filtering graph embeddings \citep{wu_learning_2021,liu_self-supervised_2022}. Another strategy encourages representation neutrality by optimizing for representations that are orthogonal to sensitive dimensions in the representation space \citep{wu_fairness-aware_2021,islam_debiasing_2021}. There are also examples of introducing neutrality by adjusting sampling schemes applied while training representations \citep{rahman_fairwalk_2019,li_fairsr_2022}, and methods for causally isolating sensitive information to specific model factors that are replaced or dropped during inference \citep{buyl_debayes_2020,frisch_co-clustering_2021}.

The framework proposed by \citet{li2023transferable} shares the most high-level similarities with this work in considering Representation Neutrality fairness optimized through adversarial methods and in being capable of recommending for unseen users. The latter is achieved by training an auxiliary mapping function to map new users to the representation space of the trained model.

\citet{flexibly_fair_vae} propose a VAE-based classification model that allows its users to specify which sensitive attributes it should be oblivious to. One of the key contributions is to isolate all sensitive user information in one part of the latent representation. The sensitive part of the latent representation is subjected to a secondary task of classifying the user's sensitive attributes, and an approximated KL-divergence measure is minimized to encourage independence between the two parts of the latent representation.

\section{Background}
\subsection{Variational Autoencoders}
The Variational Autoencoder (VAE) is a variational Bayesian model initially proposed by \citet{vae_kingma_welling} that has since seen wide application within representation learning and image generation. The vanilla VAE is posed as a graphical model where the observed data $\vx$ depends on a latent variable $\vz$. The final objective of maximizing the Evidence Lower Bound (ELBO) can be derived from minimizing the KL-divergence from the variational distribution $q(\vz|\vx)$ to the true posterior $p(\vz|\vx)$.
\begin{equation}
   \text{ELBO}(q,p) = \mathbb{E}_{q(\vz|\vx)}\left[\log p(\vx|\vz)\right] - D_{KL}\left[q(\vz|\vx)||p(\vz)\right],
\end{equation}
\noindent where $p(\vx|\vz)$ is the likelihood and $p(\vz)$ is the latent prior. Neural networks typically parameterize $p(\vx|\vz)$ and $q(\vz|\vx)$.

\citet{beta_vae} proposed the $\beta$-VAE which adds a $\beta$ factor associated with the KL-divergence term. In particular, they explore $\beta > 1$ to produce disentangled encodings, i.e., the increased focus on independence between dimensions of $\vz$ leads to semantically different concepts encoded in them. 

Similarly, \citet{factor_vae} adds an optimization term for increased disentanglement that penalizes dependency between dimensions of the latent representation.
\begin{equation}
   \text{Loss}_{\text{FactorVAE}}(q,p) = \text{ELBO}(q,p) - \gamma D_{KL}\left[q(\vz)||\prod_{i} q(\vz_i)\right] 
\end{equation}

\subsection{VAE-based Recommender System}
\citet{vae_collab} propose applying a VAE as a Recommender System by having $\vx$ represent user interaction history, e.g., items that the user has interacted with or rated. Recommendations are extracted from the fuzzy reconstruction of the user interaction history based on values assigned to items the user has not previously interacted with. 

\section{Methodology}
\begin{figure}[htb]
\centering
\includegraphics[width=0.5\textwidth]{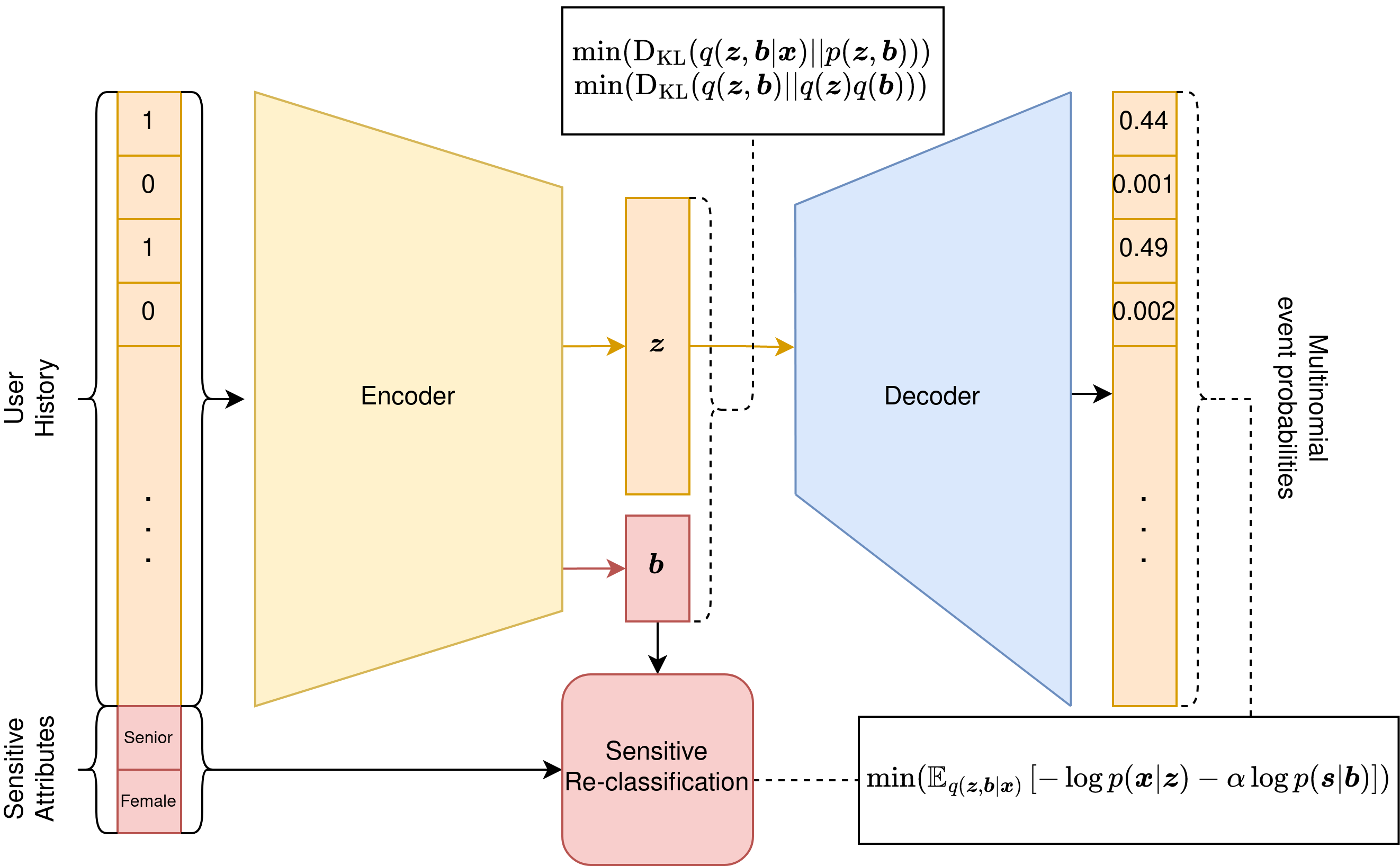}
\caption{Illustration of the Split Latent model setups. The encoder is dynamically designed and is, in practice, implemented as two separate encoders, one for $\vz$ and one for $\vb$. Key details are that no explicit sensitive information $\vs$ is provided or required during inference, and the sensitive part of the latent representation $\vb$ is not used for recommendation.}
\label{fig:model}
\end{figure}

The key idea behind our model is to leverage the bottleneck characteristics of the VAE to encourage the reduction of sensitive user information such that the provided recommendations are minimally influenced by such information. The probabilistic nature and low dimensionality of the latent representation can simplify the objective of filtering out sensitive information. The main proposed model setups are illustrated in Figure \ref{fig:model}.

\subsection{Base Model}\label{sec:base}
The underlying recommender system is based on the model proposed by \citet{vae_collab}. In addition to posing the recommendation task as the objective of reconstructing the users' interaction histories, \citet{vae_collab} propose multiple extensions. They apply dropout on the input during training (noisy VAE) to improve generalization. They explore Gaussian and Logistic likelihood $p(\vx|\vz)$ before settling on the Multinomial for its performance and nice properties. The multinomial likelihood does not explicitly penalize probability density allocated to the items the user has not interacted with, which in turn avoids the assumption of many other options in that these are all items that the user dislike and that should be allocated zero probability. 
\begin{equation}
    \log p(\vx|\vz) = \log \left[\prod_i p(\vx_i|\vz)^{\vx_i}\right]  = \sum_i \vx_i\text{log}p(\vx_i|\vz), \text{where} \sum_i p(\vx_i|\vz) = 1
\end{equation}
A specific item $\vx_i$'s contribution is zero if it is not found in the user's interaction history, regardless of what the decoder parameterizes the item's probability $p(\vx_i|\vz)$ to be. Unlike \citet{beta_vae}, \citet{vae_collab} explore $\beta < 1$, citing that generation has limited applications in the recommender system setting, and identify $\beta=0.2$ as a good candidate. 

The model of \citet{vae_collab} was altered slightly in this work: The proposed $\beta$-annealing strategy was dropped since it did not yield noticeable improvements, the Hyperbolic Tangent activation was switched out with the SELU \citep{selu}, and better results were achieved when reducing the dimensionality of the latent state. The latent dimension was set to 64 for the baseline setup and 24 for all fairness setups, as opposed to 200 in \citet{vae_collab}. Further, $\beta$ was set to 1 for the extended setups as it was found to synergize better with the fairness extensions.

\subsection{Adversarial Setup}
The adversarial setup is considered a baseline setup and couples the base model with an adversarial model tasked with classifying sensitive attributes from the latent state $\vz$. This is one of the extensions explored by \citet{borges_f2vae_2022}. Insight from the adversarial model help the main model avoid encoding sensitive information.

\subsection{Split Latent Setups}
The split latent state setups bisect the latent state into one part for encoding sensitive user information $\vb$ and another part that is free for sensitive information $\vz$. Fair recommendations are produced by decoding $\vz$, while $\vb$ is discarded during evaluation. The motivation behind the bisected latent state is to leverage the sensitive information in $\vb$ to inform the model of the information that should not be encoded in $\vz$. An inspirational split latent setup is proposed by \citet{flexibly_fair_vae}. Our approaches differ in that we consider recommendation rather than classification, we do not limit the encoding of each sensitive attribute to single dimensions in $\vb$, and we posit isotropic Gaussian priors to $\vb$.

A classification task is coupled with $\vb$ through another decoder, which will be referred to as the Sensitive Decoder, with the goal of inferring the user's sensitive attributes. In both Split Latent setups, binary sensitive attributes were considered, and the re-classification was optimized using cross-entropy. The choice of an isotropic Gaussian prior $p(\vz,\vb)$ will inherently optimize the VAE to produce independent dimensions of $[\vz\; \vb]$, but this is supplemented with an explicit term for penalizing correlation between $\vz$ and $\vb$. The full objective to be maximized is
\begin{equation}
\begin{split}
    \text{Split}&\text{LatentObj}(q,p) =\\ &\mathbb{E}_{q(\vz,\vb|\vx)}\left[\log p(\vx|\vz) + \alpha \log p(\vs|\vb)\right] - \beta D_{KL}\left[q(\vz,\vb|\vx)||p(\vz,\vb)\right] - \gamma D_{KL}\left[q(\vz,\vb)||q(\vz)q(\vb)\right],
\end{split}
\end{equation}
where $\beta$, $\alpha$ and $\gamma$ are hyperparameters for adjusting the influence of terms, $\mathbb{E}_{q(\vz|\vx)}\left[\alpha \text{log}p(\vs|\vb)\right]$ comprise the re-classification objective of sensitive attribute $\vs$ given $\vb$, and $D_{KL}\left[q(\vz,\vb)||q(\vz)q(\vb)\right]$ is the term introduced for further penalizing correlation between $\vz$ and $\vb$. The latter is defined and implemented in two different ways.

\noindent \textbf{Generative Adversarial Network (GAN) KL.}
Inspiration has been taken from \citet{factor_vae}, who proposed approximating the KL divergence from the aggregate posterior $\hat{q}(\vz) = \mathbb{E}_{\vx \in \mathcal{D}_s}\left[q(\vz|\vx)\right]$, where $\mathcal{D}_s$ is the sampled minibatch, to a factorization over each dimension using an adversarial model. This measure is approximated by an adversarial model trained to tell the original latent representations apart from factorized ones where dimensions have been shuffled across the minibatch. 
The equivalent for $q(\vz)q(\vb)$ is to shuffle $\vb$ in its entirety across the minibatch.

\noindent \textbf{Empiric KL.}
GAN approaches are often hard to optimize due to the moving objectives and the balancing of two competing models. An alternative approach is based on the analytic KL-divergence of Gaussian distributions. The empiric covariance of a minibatch of latent representations replaces the covariances of $q(\vz,\vb)$, while the covariances of $q(\vz)q(\vb)$ are replaced by the same empiric covariance matrix where all covariances between $\vz$ and $\vb$ are set to zero.
\begin{equation}
    D_{KL}\left[q(\vz,\vb)||q(\vz)q(\vb)\right] = \frac{1}{2}\left[\log \frac{|\hat{\Sigma}_2|}{|\hat{\Sigma}_1|} -d + \text{tr}\left\{\hat{\Sigma}_2^{-1} \hat{\Sigma}_1 \right\}\right], \text{when } \hat{\mu}_1 = \hat{\mu}_2,
\end{equation}
\noindent where $\hat{\mu}$ are empiric means, $\hat{\Sigma}$ are empiric covariances, $d$ is the number of dimensions in the latent state, and $\text{tr}\{\}$ is the trace operator. The covariance matrix $\hat{\Sigma}_2$ is block-diagonal, so its inverse can be computed block by block. 

\section{Results}
\subsection{Experimental Setup}
Two established datasets were used to conduct the experiments. Unlike contemporary work, both datasets were split into training, validation and test datasets containing disjoint sets of users to accommodate the setting of providing unseen users with recommendations. Common for both datasets was the choice of two binary sensitive attributes: age and gender. The datasets provide gender as `male', `female', or various forms for `missing'/`undefined'/`other', but the application of labels other than male and female was deemed too inconsistent to make out one or more additional sensitive labels. 
Age was also made binary to complement the gender attribute, and a threshold age of 35 was chosen. Users that miss either sensitive attribute are filtered out since these are required during training and evaluation.
\\\noindent \textbf{MovieLens1M}: The 1 million version of the MovieLens dataset \citep{movielens_2015} is the most applied dataset for evaluating consumer-side fairness in recommender systems. The raw dataset contains 1 million movie ratings and is the largest MovieLens dataset that provides user attributes that are considered sensitive. Following established practices for converting ratings into implicit feedback data, ratings of 4 or 5 out of 5 were labelled 1 while lower ratings were set to 0.
\\\noindent \textbf{LastFM 2B}: The LFM-2b dataset \citep{melchiorre_investigating_2021} comprises 2 billion listening events collected on the lastFM platform. This dataset was processed to focus on the most recent and relevant data, as well as reducing the number of items. The most recent two years of data were extracted, and the objective was set to recommend artists rather than albums or songs. Finally, artists with fewer than 110 listening events were filtered out. Key statistics of the datasets are summarized in Table \ref{tab:datasets}.
\begin{table}[htb]
    \centering
    \caption{Key dataset statistics.}
    \begin{tabular}{ c | c c c c c }
         \textbf{Dataset} & \textbf{\# Users} & \textbf{\#Female/\#Male} & \textbf{\#Senior/\#Young} & \textbf{\# Items} & \textbf{\# Records} \\ 
         \hline
         MovieLens & 6k & ~1.7k/4.3k & ~1.4k/4.6k & ~3.5k & ~575k \\  
         LastFM& ~14.6k & ~3k/11.5k & ~1.3k/13.3k& ~22k & ~8750k \\
         \bottomrule
    \end{tabular}
    \label{tab:datasets}
\end{table}
\subsubsection{Metrics}
\textbf{NDCG@k}: NDCG@k was used as the recommendation utility metric. NDCG is a popular ranking metric in recommender systems that rewards ranking relevant items high. For all experiments, k was set to 10.
\\\noindent\textbf{AUC}: AUC(Area Under the ROC Curve) is a metric commonly used for classification that considers the True Positive Rate and the False Positive Rate for all possible threshold values. For fair recommender systems it has been used to evaluate how well sensitive information has been filtered out of representations, which is its role in this work. For this particular objective and binary sensitive attributes, the perfect AUC score is 0.5 which indicates that all sensitive information has been filtered out. AUC is measured by training an auxiliary classification model on the model representations. One challenge with this metric is that it cannot be applied to the SLIM baseline.
\\\noindent $\boldsymbol{\chi^2}$\textbf{-statistic}: $\chi^2$-test can be used to estimate the probability that two independent samples were drawn from the same distribution. For the considered setting, it is natural to compare how often items occur in recommendations given to different sensitive groups, e.g., young and senior users. Top recommendations are not independent, rendering the test unreliable, and we instead focus on the statistic. The top 100 recommendations given to each user were selected to aggregate the contingency tables of each sensitive attribute. The number of items considered was set for each dataset such that each cell had an expected value of at least 3 in all setups since the long-tail nature of recommendation yields a lot of cells with low expectations and since the statistic is highly sensitive to such cells. Expectations are adjusted according to the number of users who can be recommended each item, i.e., users who have not already interacted with it. One issue with considering the top recommendations observations is that we lose the ordering information in the ranking, where rank 1 is assigned more confidence than rank 100.
\\\noindent\textbf{Kendall-Tau distance}: Kendall-Tau distance is a distance metric between two ordered lists. The original distance is undefined when items only occur in one of the lists, so an extension designed for the recommender setting\footnote{https://godatadriven.com/blog/using-kendalls-tau-to-compare-recommendations/} is applied. The extension is shown to have some intuitive properties and will output values from 1, being a perfect match, to -1, when the original recommendation lists contain disjointed sets of items. This metric was also considered for top-$k$ recommendations given to sensitive groups, where $k$ is set to 100. The recommendations assigned to each sensitive group were aggregated over individual user recommendations while applying a rank discounting scheme to give higher importance to highly ranked items and to serve as a shared normalization strategy for the different models. 

\subsubsection{Models}
Since no implementations of viable fair baselines have been identified, comparisons are made with non-fair baselines and focus on how the different proposed fairness extensions of a base VAE recommender system impact recommendation utility and fairness measures. All code is publically available on GitHub\footnote{https://github.com/BjornarVass/fair-vae-rec}.
\\\noindent\textbf{SLIM}: SLIM \citep{slim_2011} is an established baseline recommender system that culminates in an item-to-item parameter matrix. The item-to-item nature allows it to recommend for users not considered during model fitting.
\\\noindent\textbf{VAErec}: The model proposed by \citet{vae_collab} with minor alterations as specified in Section \ref{sec:base}. This model is not trained with any fairness objectives and serves as the base model that is extended to make out the other VAE models.
\\\noindent\textbf{VAEadv}: VAERec extended with an adversarial model that filters sensitive information from the latent representation.
\\\noindent\textbf{VAEgan}: VAERec extended with bisected latent representation for isolating and filtering out sensitive information. Independence between sections is optimized using an adversarial model that approximates a KL-divergence term.
\\\noindent\textbf{VAEemp}: VAERec extended with bisected latent representation for isolating and filtering out sensitive information. Independence between sections is optimized using an analytic KL-divergence term with empiric covariances.

\subsection{Main Results}
VAErec achieves comparable NDCG as SLIM on the MovieLens dataset, but SLIM performs better on the LastFM dataset. This is contrary to the results of \citet{vae_collab} which showed improved NDCG@100 over SLIM on two movie datasets, of which the results on the 20 million version of MovieLens were successfully reproduced using our implementation. It is unclear which factors affect the performance of the VAE-recommender, particularly since our changes further improved the original model and considering that two datasets from the same source conflict in performance.

All fairness extensions are shown to improve the AUC scores significantly. In particular, the AUC scores of VAErec on the MovieLens dataset indicate that the latent aptly encodes user gender and age with AUCs above 0.8, whereas VAEemp reduces this to 0.65-0.63 at the cost of 0.035 NDCG ($\approx 11\%$). VAEemp outperforms the other options when considering NDCG and Representation Neutrality and manages to filter out age better than gender despite the base model inherently encoding more age information. The secondary fairness metrics suggest that all fair extensions produce more similar recommendations for users of different sensitive groups than SLIM and VAErec. The $\chi^2$-statistic of VAErec is worse than that of SLIM, but the fair extensions all significantly outperform SLIM. Interestingly, VAEemp is the worst-performing fair extension on these metrics, which suggests they do not perfectly reflect the AUC metric.
\begin{table}[htb]
    \tiny
    \centering
    \caption{MovieLens results.}
    \begin{tabular}{ c c c c c c c c}
         \textbf{Model} & \textbf{NDCG@10}$\uparrow$& \textbf{AUC G}$\downarrow$ & \textbf{AUC A}$\downarrow$&  $\boldsymbol{\chi^2}$\textbf{@100 G}$\downarrow$ & $\boldsymbol{\chi^2}$\textbf{@100 A}$\downarrow$ & \textbf{K.T@100 G}$\uparrow$ & \textbf{K.T@100 A}$\uparrow$ \\ 
         \toprule
         \textbf{SLIM} & \textbf{0.328}$\pm$0.009 & - & - & 2285$\pm$280.1 & 2198$\pm$237.6 & 0.476$\pm$0.075 & 0.448$\pm$0.045 \\
         \textbf{VAErec} & 0.321$\pm$0.008 & 0.804$\pm$0.024 & 0.859$\pm$0.019 & 2990$\pm$415.9 & 2636$\pm$359.3 & 0.559$\pm$0.054 & 0.537$\pm$0.035\\
         \textbf{VAEadv} & 0.280$\pm$0.008 & 0.678$\pm$0.036 & 0.675$\pm$0.043 & 1121$\pm$273.4 & 904.6$\pm$194.5 & 0.820$\pm$0.025 & 0.792$\pm$0.038\\
         \textbf{VAEgan} & 0.277$\pm$0.010 & 0.687$\pm$0.037 & 0.695$\pm$0.050 & \textbf{1054}$\pm$232.6 & \textbf{852.5}$\pm$208.0 & \textbf{0.854}$\pm$0.036 & \textbf{0.841}$\pm$0.029\\
         \textbf{VAEemp} & 0.286$\pm$0.008 & \textbf{0.652}$\pm$0.032 & \textbf{0.629}$\pm$0.041 & 1355$\pm$302.7 & 1151$\pm$228.8 & 0.804$\pm$0.033 & 0.770$\pm$0.043 \\\bottomrule
    \end{tabular}
    \label{tab:movielens}
\end{table}

The LastFM dataset offers different insights on minority groups since the sensitive groups are very skewed, e.g., only ~10\% of the users are considered senior. This seems to be reflected in a smaller improvement of Representation Neutrality achieved by the fair extensions compared with their performance on the MovieLens dataset. The improved gender AUC is comparable, but the base AUC of VAErec is lower. All three fair extensions performed similarly on gender, but VAEemp struggles more than the other two on age. The extensions' lesser improvements of AUC over VAErec come at a smaller relative reduction in NDCG than the one seen on the Movielens dataset (< 4\% for VAEadv and VAEemp). 

Kendall Tau on LastFM is significantly better for all VAE-based models, and the fair models all outperform VAErec. On the other hand, only VAEadv achieve a $\chi^2$-statistic that is better than that of SLIM for this dataset. One confounding factor is that SLIM is observed to provide more diverse recommendations, with roughly 60\% of recommendations being the top 10\% popular items, vs roughly 85\% for VAEemp, meaning that the 750 most popular items considered in the $\chi^2$-statistic of LastFM cover far fewer of the total SLIM recommendations. When comparing the VAE-based models, it is clear that the fair extensions succeed in reducing the very large initial $\chi^2$-statistics of VAErec.
\begin{table}[htb]
    \tiny
    \centering
    \caption{LastFM results.}
    \begin{tabular}{ c c c c c c c c}
         \textbf{Model} & \textbf{NDCG@10}$\uparrow$& \textbf{AUC G}$\downarrow$ & \textbf{AUC A}$\downarrow$&  $\boldsymbol{\chi^2}$\textbf{@100 G}$\downarrow$ & $\boldsymbol{\chi^2}$\textbf{@100 A}$\downarrow$ & \textbf{K.T@100 G}$\uparrow$ & \textbf{K.T@100 A}$\uparrow$ \\ 
         \toprule
         \textbf{SLIM} & \textbf{0.572}$\pm$0.010 & - & - & 2681$\pm$481.7 & 2164$\pm$292.6 & 0.238$\pm$0.099 & -0.032$\pm$0.061\\
         \textbf{VAErec} & 0.512$\pm$0.010 & 0.759$\pm$0.014 & 0.821$\pm$0.023 & 3693$\pm$661.9 & 2840$\pm$485.3 & 0.608$\pm$0.046 & 0.440$\pm$0.073\\
         \textbf{VAEadv} & 0.492$\pm$0.010 & 0.666$\pm$0.019 & \textbf{0.731}$\pm$0.025 & \textbf{2441}$\pm$546.6 & \textbf{2068}$\pm$416.8 & 0.757$\pm$0.041 & 0.655$\pm$0.043\\
         \textbf{VAEgan} & 0.484$\pm$0.014 & 0.674$\pm$0.035 & 0.738$\pm$0.031 & 2596$\pm$656.8 & 2191$\pm$601.3 & \textbf{0.775}$\pm$0.054 & \textbf{0.683}$\pm$0.061\\
         \textbf{VAEemp} & 0.495$\pm$0.010 & \textbf{0.663}$\pm$0.022 & 0.757$\pm$0.027 & 2924$\pm$577.4 & 2528$\pm$481.4 & 0.725$\pm$0.045 & 0.588$\pm$0.046 \\\bottomrule
    \end{tabular}
    \label{tab:lastfm}
\end{table}

\subsection{Sampling Feature}
VAE-based models typically only consider the parameterized mean of the variational distribution during inference, i.e., it is deterministic. In a setting where fairness is of utmost importance, one can leverage the parameterized mean and variation to sample latent states that are inherently noisy. Sampled latent states are typically fairer, i.e., more neutral, but produce less accurate recommendations. Thus, VAE-based recommender systems can dynamically offer two different modes depending on how the user values the performance and fairness tradeoff. 

The sampling feature is compared for different values of the hyperparameter $\beta$ since it directly controls the loss term regulates the parameterized distribution. The default setting of $\beta=1.0$ yields the biggest difference in results when applying sampled or deterministic latent states. Sampling with $\beta=1.0$ resulted in the best fairness scores but also the worst NDCG score. On the other hand, $\beta$ set to 0.2 and 0.6 produced marginally better NDCG and AUC for the deterministic mode with $\beta=0.2$ coming out on top. Reducing $\beta$ appears to improve the fairness of the deterministic mode at the loss of fairness in the sampling mode. Notably, sampling with $\beta=0.2$ and not sampling with $\beta=1$ achieves similar NDCG, but the AUCs achieved by the former is noticably lower. This suggests that coupling the choice of $\beta$ and the sampling strategy may yield good settings for scenarios where one metric is assigned a strict upper or lower bound constraint. Large $\beta$ may be ideal in dynamic scenarios where users can choose to turn on sampling to improve fairness.
\begin{table}[htb]
    \small
    \centering
        \caption{Results from different $\beta$ settings, with and without sampling latent representations.}
    \begin{tabular}{ r c c c }
         \textbf{Model} & \textbf{NDCG@10}$\uparrow$& \textbf{AUC G}$\downarrow$ & \textbf{AUC A}$\downarrow$\\ 
         \toprule
         \textbf{VAEemp} $\beta=1.0$& 0.286$\pm$0.008 & 0.651$\pm$0.032 & 0.629$\pm$0.041 \\
         \textbf{sampled} & 0.256$\pm$0.009 & 0.595$\pm$0.035 & 0.562$\pm$0.026 \\
         \hline
         \textbf{VAEemp} $\beta=0.6$& 0.292$\pm$0.009 & 0.652$\pm$0.029 & 0.615$\pm$0.042 \\
         \textbf{sampled} & 0.269$\pm$0.007 & 0.603$\pm$0.036 & 0.573$\pm$0.038 \\
         \hline
         \textbf{VAEemp} $\beta=0.2$& 0.292$\pm$0.008 & 0.640$\pm$0.044 & 0.607$\pm$0.030 \\
         \textbf{sampled} & 0.279$\pm$0.008 & 0.619$\pm$0.042 & 0.587$\pm$0.038 \\
         \bottomrule
    \end{tabular}
    \label{tab:sample}
\end{table}

\section{Conclusion and Future Work}
Relating to the first research question, this research indicates that VAE-based recommenders consistently perform well but are not universally competitive. While few architectures excel in all scenarios, this insight may motivate research into other model types capable of providing new users with fair recommendations. To answer the second research question, we have shown that these models can be successfully applied and extended to significantly limit the demographic information encoded in the latent state, meaning that they can provide new and established users alike with fairer recommendations. The improvement in fairness comes with a minor deterioration of recommendation utility, as is seen in similar research. The VAE also offers a means for further obfuscating user representation through parameterized latent state variance. For the primary fairness definition of Neutral Representation, the proposed extensions significantly outperform the base model. The models also perform well on the secondary fairness definition of Recommendation Parity where the proposed VAE extensions either outperform or match relevant models on two metrics. 

For future work, it could be interesting to evaluate the model on other datasets representing different recommendation settings. It would also be desirable to compare with other models that optimize for fairness when recommending for new users. Such models could be based on contextual recommender systems or other architectures that do not learn explicit user representations.

\section*{Acknowledgments}
This publication has been partly funded by the SFI NorwAI, (Centre for Research-based Innovation, 309834). The authors gratefully acknowledge the financial support from the Research Council of Norway and the partners of the SFI NorwAI.

\bibliographystyle{plainnat}
\bibliography{references}

\begin{thebibliography}{27}
\providecommand{\natexlab}[1]{#1}
\providecommand{\url}[1]{\texttt{#1}}
\expandafter\ifx\csname urlstyle\endcsname\relax
  \providecommand{\doi}[1]{doi: #1}\else
  \providecommand{\doi}{doi: \begingroup \urlstyle{rm}\Url}\fi

\bibitem[Ahanger et~al.(2022)Ahanger, Aalam, Bhat, and Assad]{pop_bias}
Abdul~Basit Ahanger, Syed~Wajid Aalam, Muzafar~Rasool Bhat, and Assif Assad.
\newblock Popularity bias in recommender systems - a review.
\newblock In Valentina~E. Balas, G.~R. Sinha, Basant Agarwal, Tarun~Kumar
  Sharma, Pankaj Dadheech, and Mehul Mahrishi, editors, \emph{Emerging
  Technologies in Computer Engineering: Cognitive Computing and Intelligent
  IoT}, pages 431--444, Cham, 2022. Springer International Publishing.
\newblock ISBN 978-3-031-07012-9.

\bibitem[Borges and Stefanidis(2022)]{borges_f2vae_2022}
Rodrigo Borges and Kostas Stefanidis.
\newblock {F2VAE}: a framework for mitigating user unfairness in recommendation
  systems.
\newblock In \emph{Proceedings of the 37th {ACM}/{SIGAPP} {Symposium} on
  {Applied} {Computing}}, pages 1391--1398, Virtual Event, April 2022. ACM.
\newblock ISBN 978-1-4503-8713-2.
\newblock \doi{10.1145/3477314.3507152}.

\bibitem[Bose and Hamilton(2019)]{bose_compositional_2019}
Avishek Bose and William Hamilton.
\newblock Compositional {Fairness} {Constraints} for {Graph} {Embeddings}.
\newblock In Kamalika Chaudhuri and Ruslan Salakhutdinov, editors,
  \emph{Proceedings of the 36th {International} {Conference} on {Machine}
  {Learning}}, volume~97 of \emph{Proceedings of {Machine} {Learning}
  {Research}}, pages 715--724, Long Beach Convention Center, Long Beach, June
  2019. PMLR.
\newblock URL \url{https://proceedings.mlr.press/v97/bose19a.html}.

\bibitem[Burke(2017)]{burke_multisided_2017}
Robin Burke.
\newblock Multisided fairness for recommendation.
\newblock \emph{CoRR}, abs/1707.00093, 2017.
\newblock URL \url{http://arxiv.org/abs/1707.00093}.

\bibitem[Buyl and Bie(2020)]{buyl_debayes_2020}
Maarten Buyl and Tijl~De Bie.
\newblock {DeBayes}: a {Bayesian} {Method} for {Debiasing} {Network}
  {Embeddings}.
\newblock In \emph{Proceedings of the 37th {International} {Conference} on
  {Machine} {Learning}, {ICML} 2020, 13-18 {July} 2020, {Virtual} {Event}},
  volume 119 of \emph{Proceedings of {Machine} {Learning} {Research}}, pages
  1220--1229, Virtual, 2020. PMLR.
\newblock URL \url{http://proceedings.mlr.press/v119/buyl20a.html}.

\bibitem[Creager et~al.(2019)Creager, Madras, Jacobsen, Weis, Swersky, Pitassi,
  and Zemel]{flexibly_fair_vae}
Elliot Creager, David Madras, Joern-Henrik Jacobsen, Marissa Weis, Kevin
  Swersky, Toniann Pitassi, and Richard Zemel.
\newblock Flexibly fair representation learning by disentanglement.
\newblock In Kamalika Chaudhuri and Ruslan Salakhutdinov, editors,
  \emph{Proceedings of the 36th International Conference on Machine Learning},
  volume~97 of \emph{Proceedings of Machine Learning Research}, pages
  1436--1445, Long Beach Convention Center, Long Beach, 09--15 Jun 2019. PMLR.
\newblock URL \url{https://proceedings.mlr.press/v97/creager19a.html}.

\bibitem[Farnadi et~al.(2018)Farnadi, Kouki, Thompson, Srinivasan, and
  Getoor]{hybrid_farnadi}
Golnoosh Farnadi, Pigi Kouki, Spencer~K. Thompson, Sriram Srinivasan, and Lise
  Getoor.
\newblock A fairness-aware hybrid recommender system.
\newblock \emph{CoRR}, abs/1809.09030, 2018.
\newblock URL \url{http://arxiv.org/abs/1809.09030}.

\bibitem[Frisch et~al.(2021)Frisch, Leger, and
  Grandvalet]{frisch_co-clustering_2021}
Gabriel Frisch, Jean-Benoist Leger, and Yves Grandvalet.
\newblock Co-clustering for {Fair} {Recommendation}.
\newblock In Michael Kamp, Irena Koprinska, Adrien Bibal, Tassadit Bouadi, and
  Benoît Frénay, editors, \emph{Machine {Learning} and {Principles} and
  {Practice} of {Knowledge} {Discovery} in {Databases}}, pages 607--630, Cham,
  2021. Springer International Publishing.
\newblock ISBN 978-3-030-93736-2.

\bibitem[Harper and Konstan(2015)]{movielens_2015}
F.~Maxwell Harper and Joseph~A. Konstan.
\newblock The movielens datasets: History and context.
\newblock \emph{ACM Trans. Interact. Intell. Syst.}, 5\penalty0 (4), dec 2015.
\newblock ISSN 2160-6455.
\newblock \doi{10.1145/2827872}.
\newblock URL \url{https://doi.org/10.1145/2827872}.

\bibitem[Higgins et~al.(2017)Higgins, Matthey, Pal, Burgess, Glorot, Botvinick,
  Mohamed, and Lerchner]{beta_vae}
Irina Higgins, Lo{\"{\i}}c Matthey, Arka Pal, Christopher~P. Burgess, Xavier
  Glorot, Matthew~M. Botvinick, Shakir Mohamed, and Alexander Lerchner.
\newblock beta-vae: Learning basic visual concepts with a constrained
  variational framework.
\newblock In \emph{5th International Conference on Learning Representations,
  {ICLR} 2017, Toulon, France, April 24-26, 2017, Conference Track
  Proceedings}, Toulon, France, 2017. OpenReview.net.
\newblock URL \url{https://openreview.net/forum?id=Sy2fzU9gl}.

\bibitem[Islam et~al.(2021)Islam, Keya, Zeng, Pan, and
  Foulds]{islam_debiasing_2021}
Rashidul Islam, Kamrun~Naher Keya, Ziqian Zeng, Shimei Pan, and James Foulds.
\newblock Debiasing {Career} {Recommendations} with {Neural} {Fair}
  {Collaborative} {Filtering}.
\newblock In \emph{Proceedings of the {Web} {Conference} 2021}, pages
  3779--3790, Ljubljana Slovenia, April 2021. ACM.
\newblock ISBN 978-1-4503-8312-7.
\newblock \doi{10.1145/3442381.3449904}.

\bibitem[Kamishima et~al.(2018)Kamishima, Akaho, Asoh, and
  Sakuma]{kamishima_recommendation_2018}
Toshihiro Kamishima, Shotaro Akaho, Hideki Asoh, and Jun Sakuma.
\newblock Recommendation {Independence}.
\newblock In Sorelle~A. Friedler and Christo Wilson, editors, \emph{Conference
  on {Fairness}, {Accountability} and {Transparency}, {FAT} 2018, 23-24
  {February} 2018, {New} {York}, {NY}, {USA}}, volume~81 of \emph{Proceedings
  of {Machine} {Learning} {Research}}, pages 187--201, New York, USA, 2018.
  PMLR.
\newblock URL \url{http://proceedings.mlr.press/v81/kamishima18a.html}.

\bibitem[Kim and Mnih(2018)]{factor_vae}
Hyunjik Kim and Andriy Mnih.
\newblock Disentangling by factorising.
\newblock In Jennifer Dy and Andreas Krause, editors, \emph{Proceedings of the
  35th International Conference on Machine Learning}, volume~80 of
  \emph{Proceedings of Machine Learning Research}, pages 2649--2658, Stockholm,
  Sweden, 10--15 Jul 2018. PMLR.
\newblock URL \url{https://proceedings.mlr.press/v80/kim18b.html}.

\bibitem[Kingma and Welling(2014)]{vae_kingma_welling}
Diederik~P. Kingma and Max Welling.
\newblock Auto-encoding variational bayes.
\newblock In Yoshua Bengio and Yann LeCun, editors, \emph{2nd International
  Conference on Learning Representations, {ICLR} 2014, Banff, AB, Canada, April
  14-16, 2014, Conference Track Proceedings}, Banff, AB, Canada, 2014. ICLR.
\newblock URL \url{http://arxiv.org/abs/1312.6114}.

\bibitem[Klambauer et~al.(2017)Klambauer, Unterthiner, Mayr, and
  Hochreiter]{selu}
G\"{u}nter Klambauer, Thomas Unterthiner, Andreas Mayr, and Sepp Hochreiter.
\newblock Self-normalizing neural networks.
\newblock In I.~Guyon, U.~Von Luxburg, S.~Bengio, H.~Wallach, R.~Fergus,
  S.~Vishwanathan, and R.~Garnett, editors, \emph{Advances in Neural
  Information Processing Systems}, volume~30, Montreal, Canada, 2017. Curran
  Associates, Inc.
\newblock URL
  \url{https://proceedings.neurips.cc/paper_files/paper/2017/file/5d44ee6f2c3f71b73125876103c8f6c4-Paper.pdf}.

\bibitem[Li et~al.(2022)Li, Hsu, and Zhang]{li_fairsr_2022}
Cheng-Te Li, Cheng Hsu, and Yang Zhang.
\newblock Fairsr: Fairness-aware sequential recommendation through multi-task
  learning with preference graph embeddings.
\newblock \emph{ACM Trans. Intell. Syst. Technol.}, 13\penalty0 (1), feb 2022.
\newblock ISSN 2157-6904.
\newblock \doi{10.1145/3495163}.

\bibitem[Li et~al.(2023)Li, Wang, Chen, and Zhang]{li2023transferable}
Yunqi Li, Dingxian Wang, Hanxiong Chen, and Yongfeng Zhang.
\newblock Transferable fairness for cold-start recommendation, 2023.

\bibitem[Liang et~al.(2018)Liang, Krishnan, Hoffman, and Jebara]{vae_collab}
Dawen Liang, Rahul~G. Krishnan, Matthew~D. Hoffman, and Tony Jebara.
\newblock Variational autoencoders for collaborative filtering.
\newblock In \emph{Proceedings of the 2018 World Wide Web Conference}, WWW '18,
  page 689–698, Republic and Canton of Geneva, CHE, 2018. International World
  Wide Web Conferences Steering Committee.
\newblock ISBN 9781450356398.
\newblock \doi{10.1145/3178876.3186150}.
\newblock URL \url{https://doi.org/10.1145/3178876.3186150}.

\bibitem[Liu et~al.(2022)Liu, Lin, Fan, Ren, Xu, Zhang, Wen, Zhao, Lin, and
  Yang]{liu_self-supervised_2022}
Haifeng Liu, Hongfei Lin, Wenqi Fan, Yuqi Ren, Bo~Xu, Xiaokun Zhang, Dongzhen
  Wen, Nan Zhao, Yuan Lin, and Liang Yang.
\newblock Self-supervised learning for fair recommender systems.
\newblock \emph{Applied Soft Computing}, 125:\penalty0 109126, August 2022.
\newblock ISSN 1568-4946.
\newblock \doi{10.1016/j.asoc.2022.109126}.

\bibitem[Melchiorre et~al.(2021)Melchiorre, Rekabsaz, Parada-Cabaleiro, Brandl,
  Lesota, and Schedl]{melchiorre_investigating_2021}
Alessandro~B. Melchiorre, Navid Rekabsaz, Emilia Parada-Cabaleiro, Stefan
  Brandl, Oleg Lesota, and Markus Schedl.
\newblock Investigating gender fairness of recommendation algorithms in the
  music domain.
\newblock \emph{Information Processing \& Management}, 58\penalty0
  (5):\penalty0 102666, 2021.
\newblock ISSN 0306-4573.
\newblock \doi{https://doi.org/10.1016/j.ipm.2021.102666}.
\newblock URL
  \url{https://www.sciencedirect.com/science/article/pii/S0306457321001540}.

\bibitem[Ning and Karypis(2011)]{slim_2011}
Xia Ning and George Karypis.
\newblock {SLIM:} sparse linear methods for top-n recommender systems.
\newblock In Diane~J. Cook, Jian Pei, Wei Wang, Osmar~R. Za{\"{\i}}ane, and
  Xindong Wu, editors, \emph{11th {IEEE} International Conference on Data
  Mining, {ICDM} 2011, Vancouver, BC, Canada, December 11-14, 2011}, pages
  497--506, Vancouver, Canada, 2011. {IEEE} Computer Society.
\newblock \doi{10.1109/ICDM.2011.134}.
\newblock URL \url{https://doi.org/10.1109/ICDM.2011.134}.

\bibitem[Rahman et~al.(2019)Rahman, Surma, Backes, and
  Zhang]{rahman_fairwalk_2019}
Tahleen Rahman, Bartlomiej Surma, Michael Backes, and Yang Zhang.
\newblock Fairwalk: {Towards} {Fair} {Graph} {Embedding}.
\newblock In \emph{Proceedings of the {Twenty}-{Eighth} {International} {Joint}
  {Conference} on {Artificial} {Intelligence}}, pages 3289--3295, Macao, China,
  August 2019. International Joint Conferences on Artificial Intelligence
  Organization.
\newblock ISBN 978-0-9992411-4-1.
\newblock \doi{10.24963/ijcai.2019/456}.

\bibitem[Resheff et~al.(2019)Resheff, Elazar, Shahar, and
  Shalom]{resheff_privacy_2019}
Yehezkel~S. Resheff, Yanai Elazar, Moni Shahar, and Oren~Sar Shalom.
\newblock Privacy and {Fairness} in {Recommender} {Systems} via {Adversarial}
  {Training} of {User} {Representations}.
\newblock In Maria~De Marsico, Gabriella Sanniti~di Baja, and Ana L.~N. Fred,
  editors, \emph{Proceedings of the 8th {International} {Conference} on
  {Pattern} {Recognition} {Applications} and {Methods}, {ICPRAM} 2019,
  {Prague}, {Czech} {Republic}, {February} 19-21, 2019}, pages 476--482,
  Prague, Czech Republic, 2019. SciTePress.
\newblock \doi{10.5220/0007361204760482}.

\bibitem[Wu et~al.(2021{\natexlab{a}})Wu, Wu, Wang, Huang, and
  Xie]{wu_fairness-aware_2021}
Chuhan Wu, Fangzhao Wu, Xiting Wang, Yongfeng Huang, and Xing Xie.
\newblock Fairness-aware news recommendation with decomposed adversarial
  learning.
\newblock In \emph{Thirty-Fifth {AAAI} Conference on Artificial Intelligence,
  {AAAI} 2021, Thirty-Third Conference on Innovative Applications of Artificial
  Intelligence, {IAAI} 2021, The Eleventh Symposium on Educational Advances in
  Artificial Intelligence, {EAAI} 2021, Virtual Event, February 2-9, 2021},
  pages 4462--4469, Virtual, 2021{\natexlab{a}}. {AAAI} Press.
\newblock URL \url{https://ojs.aaai.org/index.php/AAAI/article/view/16573}.

\bibitem[Wu et~al.(2021{\natexlab{b}})Wu, Chen, Shao, Hong, Wang, and
  Wang]{wu_learning_2021}
Le~Wu, Lei Chen, Pengyang Shao, Richang Hong, Xiting Wang, and Meng Wang.
\newblock Learning {Fair} {Representations} for {Recommendation}: {A}
  {Graph}-{Based} {Perspective}.
\newblock In \emph{Proceedings of the {Web} {Conference} 2021}, {WWW} '21,
  pages 2198--2208, New York, NY, USA, 2021{\natexlab{b}}. Association for
  Computing Machinery.
\newblock ISBN 978-1-4503-8312-7.
\newblock \doi{10.1145/3442381.3450015}.
\newblock event-place: Ljubljana, Slovenia.

\bibitem[Xu et~al.(2021)Xu, Cui, Sun, Deng, and Zheng]{xu_fair_2021}
Bingke Xu, Yue Cui, Zipeng Sun, Liwei Deng, and Kai Zheng.
\newblock Fair {Representation} {Learning} in {Knowledge}-aware
  {Recommendation}.
\newblock In \emph{2021 {IEEE} {International} {Conference} on {Big}
  {Knowledge} ({ICBK})}, pages 385--392, Auckland, New Zealand, 2021. IEEE.
\newblock \doi{10.1109/ICKG52313.2021.00058}.

\bibitem[Yao and Huang(2017)]{beyond_2017}
Sirui Yao and Bert Huang.
\newblock Beyond parity: Fairness objectives for collaborative filtering.
\newblock In Isabelle Guyon, Ulrike von Luxburg, Samy Bengio, Hanna~M. Wallach,
  Rob Fergus, S.~V.~N. Vishwanathan, and Roman Garnett, editors, \emph{Advances
  in Neural Information Processing Systems 30: Annual Conference on Neural
  Information Processing Systems 2017, December 4-9, 2017, Long Beach, CA,
  {USA}}, pages 2921--2930, Long Beach, CA, USA, 2017. NeurIPs.
\newblock URL
  \url{https://proceedings.neurips.cc/paper/2017/hash/e6384711491713d29bc63fc5eeb5ba4f-Abstract.html}.

\end{thebibliography}

\end{document}